\documentclass[aas_macros,useAMS,usenatbib]{mn2e}
\usepackage{natbib}
\usepackage{amssymb}
\usepackage{graphicx}
\usepackage{hyperref}
\usepackage{amsmath}
\usepackage{multicol}
\usepackage{placeins}
\usepackage{deluxetable}
\newcommand{\kms}{km\,s$^{-1}$}

\title[High-e DEBs in ACVS and MACC]{The Highly-Eccentric Detached Eclipsing Binaries in ACVS and MACC}
\author[Shivvers et. al]{Isaac Shivvers$^{\dag,1}$, Joshua S. Bloom$^1$, Joseph W. Richards$^1$ \\
$^1$Astronomy Department, University of California, Berkeley, CA 94720 \\
$^\dag$Email: ishivvers@astro.berkeley.edu}
\date{Accepted for publication in MNRAS; 2014 March 25.}

\begin{document}
\maketitle

\begin{abstract}
Next-generation synoptic photometric surveys will yield unprecedented (for the astronomical community) volumes of data and the processes of discovery and rare-object identification are, by necessity, becoming more autonomous.  Such autonomous searches can be used to find objects of interest applicable to a wide range of outstanding problems in astronomy, and in this paper we present the methods and results of a largely autonomous search for highly eccentric detached eclipsing binary systems in the Machine-learned ASAS Classification Catalog. 106 detached eclipsing binaries with eccentricities of $e \gtrsim 0.1$ are presented, most of which are identified here for the first time.  We also present new radial-velocity curves and absolute parameters for 6 of those systems with the long-term goal of increasing the number of highly eccentric systems with orbital solutions, thereby facilitating further studies of the tidal circularization process in binary stars.

\end{abstract}


\section{Introduction}
Detached eclipsing binaries (DEBs) have long served as powerful astronomical tools, allowing astronomers to derive the fundamental parameters of distant stars 
\citep[e.g.][]{kopal59,popper80,gimenez85,anderson91,Torres}.
The most important results of these binary star studies are facilitated by the assumption that stars in non-interacting (`detached') orbits evolve as if they were single stars, an assumption which allows us to map our empirical understanding of detached binary systems directly onto single stars everywhere.  This assumption generally appears to be robust, but there are several outstanding problems in our understanding of the evolutionary differences between stars in binary pairs and those that have evolved singly.  One of the most important gaps in our understanding has to do with the history of orbital circularization in binary stars and just how important each star's individual tidal interaction history is.

Many binary stars display orbital eccentricities very near zero.  The primordial distribution of eccentricities formed from binary star formation is only poorly understood, but it is expected to be much wider than that exhibited in evolved stars.  The low eccentricities of evolved binaries is likely a result of tidal interactions between the two stars conspiring to circularize the orbit \citep[e.g.][and citations within]{mazeh}.  To first order, any primordial eccentricity is expected to decay exponentially on a characteristic timescale ($t_{circ}$) which depends strongly upon internal structure and orbital separation  \citep[tidal circularization is a complex topic that has featured in astrophysical discussions for many years; e.g.][]{darwin}.  Several different processes likely contribute to this decay, including dynamical and equilibrium tide effects \citep[e.g.][]{zahn75, zahn77, hut} and hydrodynamical effects \citep[e.g.][]{tassoul88, tassoul96}, depending on the masses, sizes, structure and separation of the binary components.  Final estimates of the circularization timescale can vary widely \citep[e.g.][]{zahn89}, and analyses of systems in the literature indicate problems with our understanding \citep[e.g.][]{claret1, claret2, north, meibom}. However, it is abundantly clear that most binary stars undergo some variation of circularization for some amount of their evolutionary history.  This raises an important question: how does this process affect the binary components themselves, and could it introduce a systematic difference between the properties of stars in detached binaries and singly-evolved stars, thereby affecting our measurements of the Mass-Radii or Mass-Luminosity relations for all stars?

To answer this question empirically we set out to identify and characterize binary stars in highly-eccentric orbits.  Whether because they are young or because their $t_{circ}$ is very large, these systems have (on average) undergone less of the circularization process than their $e = 0.0$ counterparts, and any systematic change wrought by that process may be measurable by comparing the properties of binary stars in high-$e$ orbits to those in circular orbits.  With the advent of very large photometric data sets and new techniques for classification the number of known DEBs has skyrocketed \citep[e.g.][]{anderson91,debil,Torres,MACC}, but the number of those that have well-measured masses and radii is quite small.  A recent review by \citet{Torres} presented the currently-known DEB systems with physical system parameters determined to an accuracy of $\pm 3 \%$ or better, of which only $15$ have $e \gtrsim 0.1$.   Radial velocity (RV) curves are required but they are time-intensive and expensive to produce and most of the high-$e$ systems found by modern surveys are relatively faint, making it difficult to obtain the high-resolution spectra necessary for accurate RVs.

However, the All-Sky Automated Survey Catalog of Variable Stars (ASAS/ACVS) photometric database probes magnitude ranges amenable to RV followup, has a long baseline of observations with a relatively high cadence (producing the well-sampled light curves needed to identify and characterize eclipsing binaries) and was recently re-analyzed with all sources re-classified through modern machine-learning techniques by \citet{MACC}, yielding the Machine-learned ASAS Classification Catalog (MACC).
DEBs are identified within MACC, though their degree of orbital eccentricity is not.  Helpfully, the techniques used by MACC enable targeted searches for anomalous sources and for rare classes of objects.  

We performed a search for highly-eccentric DEBs ($e \gtrsim 0.1$) in the MACC, yielding 106 bright (V$<$12\,mag) systems, most of which are identified here for the first time.  We also present new high-resolution RV curves with modeled physical parameters for six of these objects.  In \S \ref{sec:data} we describe our photometric and spectroscopic data; in \S \ref{sec:finding} we describe our search for eccentric eclipsing binaries and present the highly-eccentric DEBs in MACC; in \S \ref{sec:RV} we present the new RV curves and physical parameters for six systems; in \S \ref{sec:conclusion} we discuss our results.

\section{Data}
\label{sec:data}
\subsection{Photometry: ASAS and MACC}

This study utilized the Machine-learned ASAS Classification Catalog presented by \citet[MACC;][]{MACC}, which provides a rich and publicly-available data set of probabilistic classifications for objects in the All-Sky Automated Survey Catalog of Variable Stars \citep[ACVS;][]{ACVS}.  MACC is a 28--class catalog of the 50,124 sources in the ACVS.  MACC achieves an estimated classification error rate of less than $20\%$ and is the first variable star catalog to report calibrated posterior class probabilities.  
MACC includes three eclipsing binary star classes -- Beta Persei, Beta Lyrae and W Ursae Majoris. In the naming scheme of 
the General Catalog of Variable Stars \citep[GCVS,][]{GCVS}, these correspond to the EA, EB, and EW classes, respectively.
Beta Persei systems exhibit spherical (fully detached) or nearly-spherical components while
Beta Lyrae systems have strongly ellipsoidal components and W Ursae Majoris systems have strongly ellipsoidal
components at very small seperations (which may even be in contact).
This paper is concerned with the fully-detached Beta Persei systems, hereafter called detached eclipsing binaries (DEBs).
We accessed MACC through the web interface\footnote{\href{http://www.bigmacc.info/}{http://www.bigmacc.info}}, pre-selecting all 2454 objects identified as Beta Persei systems for our search.  We obtained the complete ASAS light curves for each of these sources from the DotAstro.org light-curve warehouse\footnote{\href{http://www.dotastro.org/}{http://www.dotastro.org}}.

\subsection{Spectroscopy: CHIRON and the Hamilton Spectrometer}

As part of this project, we took a set of high-resolution spectra of six systems to measure their RV curves and model the properties of their component stars.  Systems were chosen for RV followup based on their estimated eccentricity and brightness as well as telescope availability.  All six of the systems with RV curves presented here are double-lined spectroscopic binaries.
Spectra for ASAS objects 205642+1153.0 and 193043-0615.6 were obtained with the Hamilton Echelle Spectrograph \citep[HES;][]{hamspec} mounted at the Coud\'{e} focus of Lick Observatory's Shane 3\,m telescope.  The spectra were taken on successive nights and spaced more or less evenly in phase, covering approximately two periods for 205642+1153.0 and one period for 193043-0615.6.  Note that the iodine cell usually installed in the HES was out of commission (and out of the optical path) during these observations, and so characteristic RV errors obtained here are significantly larger than those achieved with other studies on the same instrument.  Otherwise, the raw cross-dispersed echelle spectra were obtained in the standard way \citep[described in detail in][]{iodine}, and reduced using standard methods \citep{echelle_reduction}.

Spectra for ASAS objects 091704-5454.1, 073611-3123.4, 064057-2637.6 and 112145-0850.2 were obtained with the CHIRON spectrograph \citep{chiron} mounted on the Small and Medium Aperture Research Telescope System (SMARTS) 1.5\,m telescope at the Cerro-Tololo Inter-American Observatory (CTIO).  The spectra were taken as part of a queue-scheduled program and were spread pseudo-randomly over the course of the Fall 2012 semester with the goal of obtaining an approximately uniform phase sampling of RV measurements for each object.  The spectra were reduced in the standard way \citep{chiron}.  Table~\ref{tab:spectra}
provides a log of all spectroscopic observations used by this study.

\section{Identifying Highly-Eccentric Binaries}
\label{sec:finding}

MACC offers several data products including the predicted class, the normalized probability of membership in each class for each object and an `Anomaly' score, $A$.  The anomaly score is a measure of the difference between a source's properties and the set of properties defined by the catalog's training set for the predicted class.  Note that there are valuable differences between a low class probability (i.e.\ a source may be a good match to several classes) and a high anomaly score (i.e.\ a source is dissimilar to any class).  The combination of a high class probability and a high anomaly score implies that a source is best-described by only a single class, yet  is also significantly dissimilar from the examples of that class in the training data.  The training set of Beta Persei stars in MACC is clustered around eccentricities of $e \sim 0$.  DEB light curves are, however, strongly dissimilar from those of other classes, and the differences between eccentric and non-eccentric DEBs are not likely to place a highly-eccentric DEB into another class.  Yet those differences are captured by the features used by MACC and therefore should be reflected in the anomaly score.  In other words, most high-$e$ DEBs in MACC will be classified as a Beta Persei system, yet will also have a high anomaly score (while most low-$e$ DEBs will have a low anomaly score).

To detect true outliers, \citet{MACC} recommend a threshold of $A \gtrsim 10$.  We adopt a gentler cut of $A > 1.0$, designed to exclude only the truly normal DEBs.  We also cut on sources determined to be in the `Beta Persei' class with $P\_Class > 0.5$, discarding sources unlikely to be DEBs.   These cuts produced a set of 2146 sources, most of which were low-$e$ DEBs, and so we developed an additional parameter to identify eccentricity.  Due to the large number of sources we decided that a robust and autonomous method of identifying high-$e$ systems was much preferred over visual light-curve inspection.  Looking forward, as catalogs continue to grow exponentially, highly-autonomous methods must become the standard.  As of yet, creating a complete model for each possible system of interest by fitting physical models -- using codes like DEBiL or JKTEBOP, for example \citep{debil, jktebop} -- is very CPU-intensive.  Instead, for this work we chose a data-driven approach which seeks to identify a few quickly-calculated parameters for each light curve without relying on detailed fits to physical models.

The first step of our search was to correctly identify the period of variability in the light curve --  period misidentification proved to be a persistent source of confusion, and in the course of this work we developed a robust method for identifying the periods of eccentric DEBs (described in \S \ref{subsec:pfind}).
After correctly identifying the period, we measure the phase difference between minima for the two most significant dips apparent in each phase-folded light-curve.  To do this, we create a model by smoothing each phase-folded light-curve with a Gaussian kernel \citep{wasserman}:
\begin{gather*}
\hat{r}(\phi) = \sum^n_{i=1}Y_i l_i(\phi)   \hspace{15 mm}
l_i(\phi) = \frac{K\left( \frac{\Phi_i - \phi}{h}\right)}{\sum^n_{i=1}K\left( \frac{\Phi_i - \phi}{h}\right)}  \\ 
K(x) = \frac{1}{\sqrt{2 \pi}} e^{-x^2/2}
\end{gather*}
where $\hat{r}(\phi)$ is the value of the model at $\phi$ for data $Y_i$ over $\Phi_i$, smoothed by the kernel $K$ with bandwidth $h$.  To avoid over- or under-smoothing, we choose an optimal value of $h$ by minimizing the Generalized Cross-Validation parameter (see \S \ref{subsec:pfind} for more details).

\begin{figure}
\includegraphics[width=\linewidth]{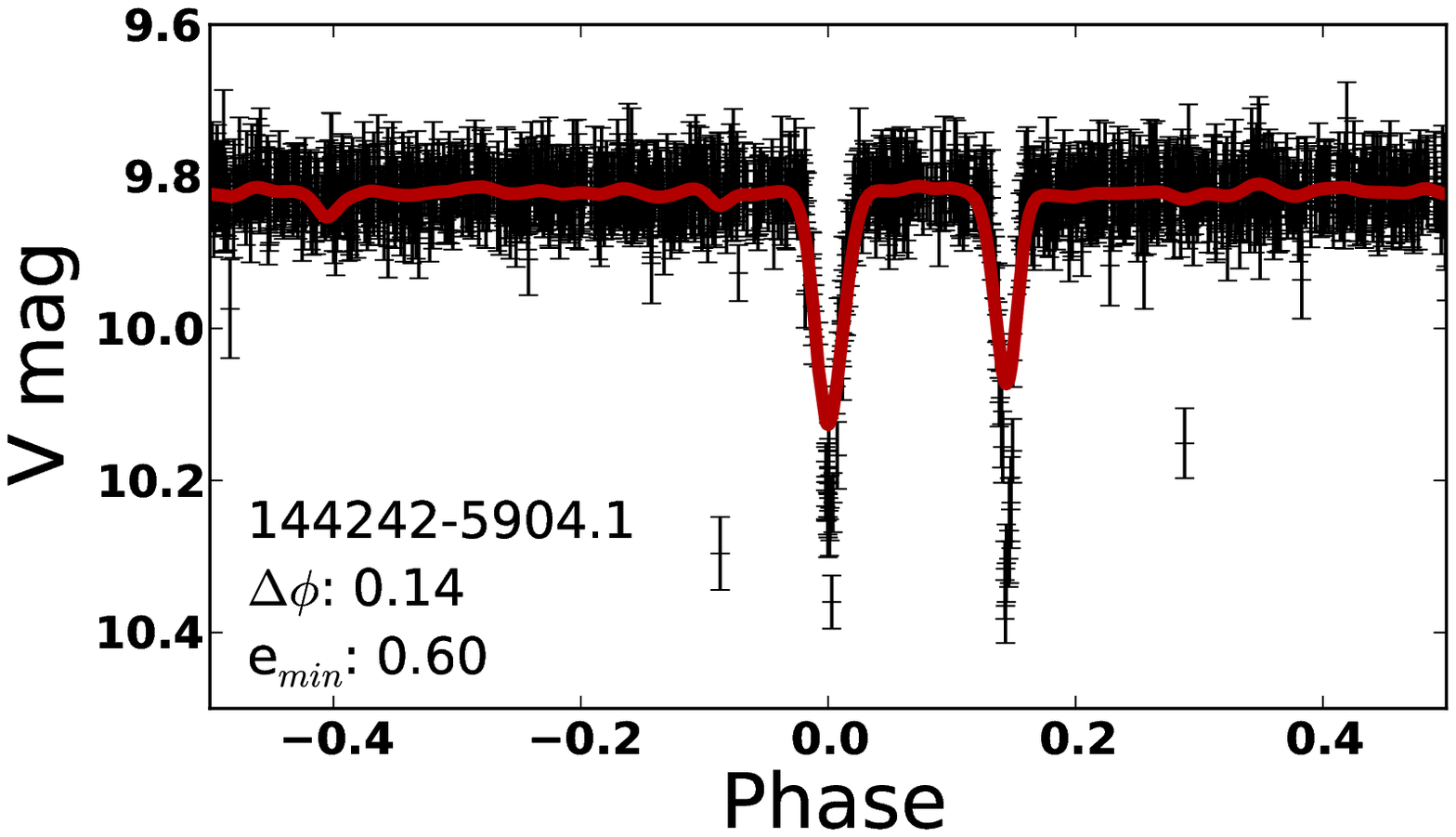}
\caption{Example ASAS lightcurve with the optimally-smoothed model overlain in red, with a smoothing kernel chosen by the GCV parameter (see \S\ref{sec:finding}).
                ASAS ID is printed in the lower left, as well as the $\Delta \phi$ and $e_{min}$ measured
                from the smoothed model.  As shown in Figure \ref{fig:array} and Table~ \ref{tab:everything},
                this object (ASAS 144242-5904.1) has a best-fit eccentricity of $e \approx 0.64$, slightly higher than the $e_{min} \approx 0.60$
                measured through the method described above.\label{fig:smooth}}
\end{figure}

This yields a smooth model with minima well-aligned in phase with the true eclipse minima (though the smoothed eclipses are generally shallower).  We record the phase difference between the lowest value in each of the two deepest smoothed minima for each light curve.  In much the same manner as \citet{dong}, we use this phase difference to establish a lower limit on the system eccentricity:
\begin{gather*}
\Delta \phi (e, \omega) = \hspace{65 mm} \\
  \frac{1}{\pi} \left[ \arccos \left(\frac{e \cos \omega}{\sqrt{1 - e^2 \sin^2 \omega}}\right) - \sqrt{1-e^2} \left( \frac{e \cos \omega}{1 - e^2 \sin^2 \omega}\right)\right]
\end{gather*}
We map the observable $\Delta \phi$ onto a lower limit for systemic eccentricity by conservatively assuming an argument of periastron of $\omega = 0$:
\begin{gather*}
\Delta \phi(e_{min}) = \frac{1}{\pi}\left(\arccos e_{min} - e_{min} \sqrt{1 - e_{min}^2}\right)
\end{gather*}
See Figure \ref{fig:smooth} for an example of an optimally-smoothed model and the measured values of $\Delta \phi$ and $e$.

\begin{figure*}
\includegraphics[width=\linewidth]{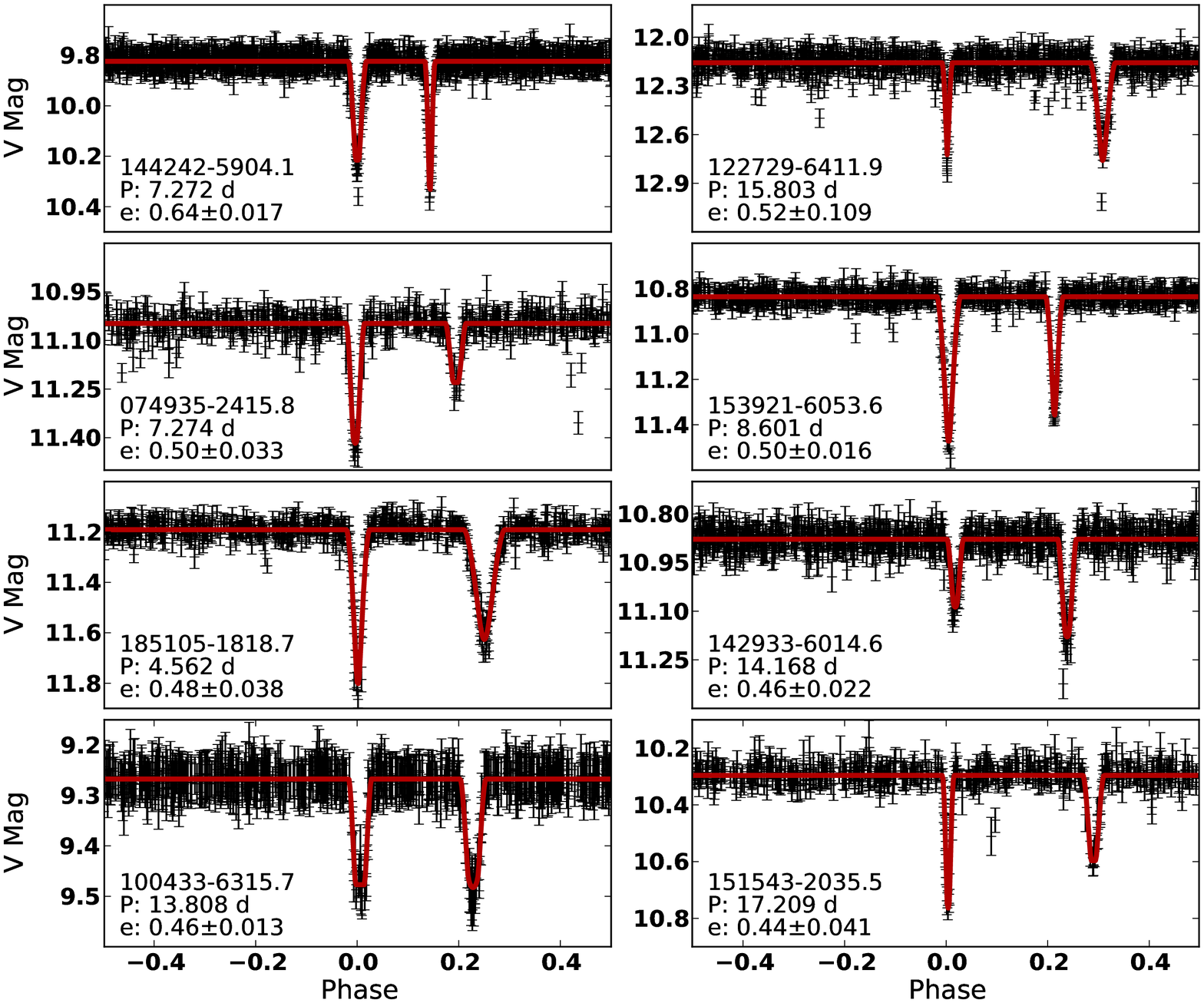}
\caption{ASAS light-curves, including reported errors, of the 8 highest-$e$ systems in Table~\ref{tab:everything}
                with model DEBiL fits overlain in red.  ASAS ID is printed in the lower left, along with period ($P$)
                and eccentricity ($e$).\label{fig:array}}
\end{figure*}

\begin{figure}
\centering
\includegraphics[width=\linewidth]{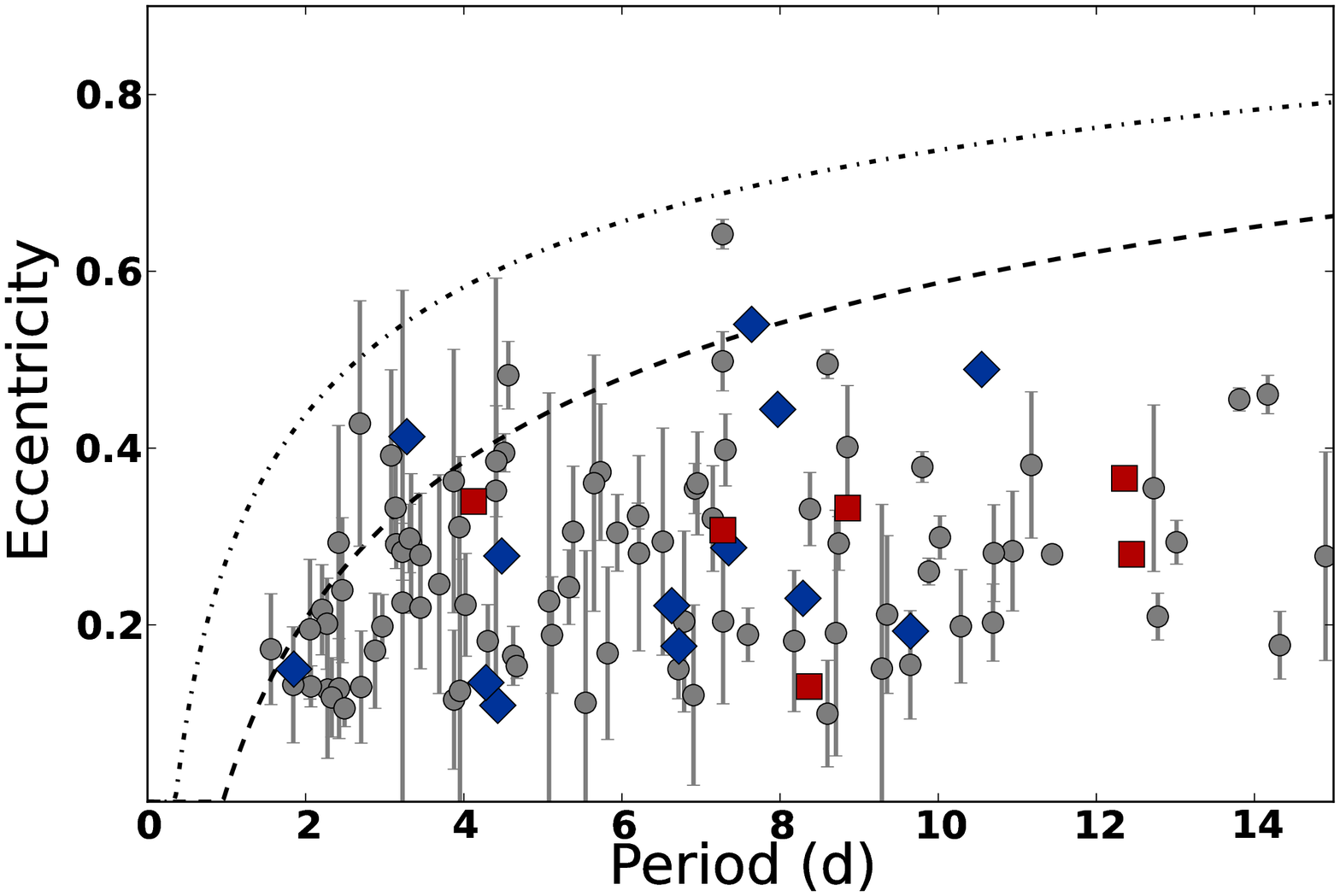}
\caption{Eccentricity and orbital period of the systems presented in Table~\ref{tab:everything} 
               and \citet{Torres}.
               Grey (circle): light curve fit with DEBiL; red (square): light curve and RV curve fit with PHOEBE; 
               blue (diamond): $e>0.1$ systems from \citet{Torres}.  Note that the errorbars are smaller than the plotted points 
               for the latter two categories.  Two curves derived from the SB9 spectroscopic catalog by \citet{mazeh} are 
               overplotted -- most binaries fall below the lower curve, and almost every known binary falls below the upper curve.  
               They both are of the form $f(P) = E - A \times e^{ - (P\times B)^c}$.  For the upper curve, $E = 0.98$, $A = 3.25$, $B = 6.3$ 
               and $C = 0.23$. For the lower,  $A = 3.5$ and $B = 3$.\label{fig:eccents}}
\end{figure}

We applied the method above to the 2146 sources chosen from MACC, selecting systems with $e_{min} > 0.10$.  This yielded a set of 161 systems which we then inspected visually.  90 of those proved to be high-$e$ DEBs.  The remainder appeared to be either DEBs with $e \approx 0$ and very noisy light-curves or mis-classified sources (not DEBs).  We then expanded our search to include all 2454 Beta Persei systems listed in MACC (keeping the cut on $e_{min}$ but no cuts on anomaly or class probability).  This yielded 78 additional systems for visual inspection, 13 of which proved to be DEBs erroneously cut from our earlier investigation.  We also include an additional 3 sources serendipitously identified elsewhere, for which the $e$-constraining method described above failed due to high noise levels.  In the end our initial cuts on reported anomaly and class probability obtained a purity of $56\%$ when coupled with the $e$-constraint described above while using the $e$-constraint alone yielded a purity of $43\%$.

In Table~\ref{tab:everything} 
we present $106$ DEBs with $e > 0.1$ including their orbital period and the best-fit eccentricity calculated using the Detached Eclipsing Binary Light curve fitting code \citep[DEBiL;][]{debil} or, for the six sources with new RV curves, the best-fit eccentricity as calculated with the Physics Of Eclipsing BinariEs  code \citep[PHOEBE;][see \S\ref{sec:modeling}]{phoebe}.  Figure \ref{fig:array} shows a few of the highest-$e$ light curves and their corresponding best-fit DEBiL models, and Figure \ref{fig:eccents} places the eccentricities of these systems in context, comparing them to other well-studied DEBs from the literature.

\subsection{Period Identification}
\label{subsec:pfind}

While examining the light curves from MACC we found that the period-finding methodology of the catalog frequently mis-estimated the period of eccentric DEBs.   The periods presented in MACC are determined through a period-finding algorithm based upon the Lomb-Scargle periodogram \citep[LS;][]{lomb,scargle,MACC}.  This method, while quite robust for most periodically-varying objects, regularly fails to identify the best period of eccentric DEBs -- the $1^{st}$-derivative discontinuities of eclipse edges and the inherent asymmetry of eclipse phases in eccentric systems conspire to limit the utility of Fourier decomposition.
This period-identification problem is certainly not a new one and several techniques have previously been developed to address it, including phase-dispersion minimization \citep{pdm, plavchan} and box-fitting least-squares \citep{bls}.  However, these algorithms tend to become incredibly CPU-intensive when processing large sets of well-sampled light-curves.  In addition, these techniques generally require a complete test of the period parameter space (as compared to a sparse minimization routine), as they do not vary smoothly over large ranges in any period-quality metric.

We found that a hierarchical combination of two approaches was highly effective.  Though the LS periodogram is unreliable at uniquely identifying the true period it does usually report a high-significance peak at the true period and/or one of its first few harmonics.  We use a generalization of the traditional LS periodogram -- described in \citet{Richards2011} -- to pre-select the five periods with the highest significance.  We add to this list of trial periods the harmonics of each period, at ratios of $2$, $3$, $\frac{1}{2}$, and $\frac{1}{3}$, to construct a set of 25 trial periods. For each trial period we fold the light-curve and perform a moving-average smooth.  We choose a smoothing bandwidth that minimizes the Generalized Cross-Validation (GCV) criterion, which penalizes over-fitting \citep{wasserman}:
\begin{gather*}
GCV(h) = \frac{1}{n}\sum^{n}_{i=1}\Bigg(\frac{Y_i - \hat{r}(\phi_i)}{1 - \frac{1}{h}}\Bigg)^2
\end{gather*}
where $h$ is the number of elements in each moving-average window, $n$ is the total number of data points, $Y$ and $\phi$ are the values and phases of the data, and $\hat{r}(\phi)$ is the smoothed model.  Note that, for a sufficiently well-sampled light-curve, choosing a value for $h$ is equivalent to choosing a width in phase over which to average the data. Finally, we select the period with the lowest GCV value.  In effect, this minimizes the GCV across both the smoothing parameter $h$ and the period, selecting for periods at which the folded data are best described by an optimally-smoothed non-parametric model.

Of the 106 eccentric DEBs presented in this paper this algorithm successfully identifies the period of 105, whereas only 19 have correct periods in MACC (`true' period determined by visual inspection).  For a randomly-selected sample of $25$ low-$e$ DEBs in MACC, our algorithm correctly identified the period of 24 and MACC correctly identified the period of 20.  Errors in period estimation are notoriously difficult to assess.  For this work, we adopt the width of the LS periodogram peak (as identified by the algorithm described above) as an estimate of our error -- characteristic errors on the periods presented in Table~\ref{tab:everything} 
are about $\pm 0.006$\,d.

\section{New RV Curves for Six High-$e$ DEBs}
\label{sec:RV}

We determined RV values through spectral cross-correlation between each spectrum and a spectrum of a radial-velocity standard star taken with the same instrument \citep[e.g.,][]{simkin74,tonry79}.  We chose not to implement a full two-dimensional cross-correlation \citep[TODCOR][]{zucker94} -- the strong similarities between eclipse depths and widths in the ASAS light curves indicated that the binary components are of similar types for these systems.  In addition, most of our spectra did not have a sufficiently high signal-to-noise ratio to attempt spectral disentangling and typing.

The HES produces a spectrum with $89$ orders, each with a wavelength range of $\sim$100\,\AA.  The reddest orders of our observations were flawed by severe fringing and so we were forced to discard all data with $\lambda > 6368$\,\AA.  
CHIRON produces a spectrum with 59 orders, each with a wavelength range of $\sim$75\,\AA.  To maximize our signal-to-noise ratio we considered only CHIRON data with $\lambda < 6862$\,\AA, where the bulk of identifiable stellar absorption lines fall.  For all observations we split each order into an integer number of equally-sized (in $\lambda$) chunks, choosing the number so as to get each chunk as close to 50\,\AA~as possible.  This chunk size parameter, as well as those parameters used in later steps of the pipeline (described below), were simply chosen to have reasonable values when considering computational expense, the density of stellar absoprtion lines and signal-to-noise ratios of our spectra.  Most orders were therefore split into two chunks, after which we removed a splined continuum fit, subtracted the standard star's RV and applied barycentric corrections to both the standard and the science star to remove the effect of Earth's orbit on the observed RVs.

\begin{figure}
\centering
\includegraphics[width=\linewidth]{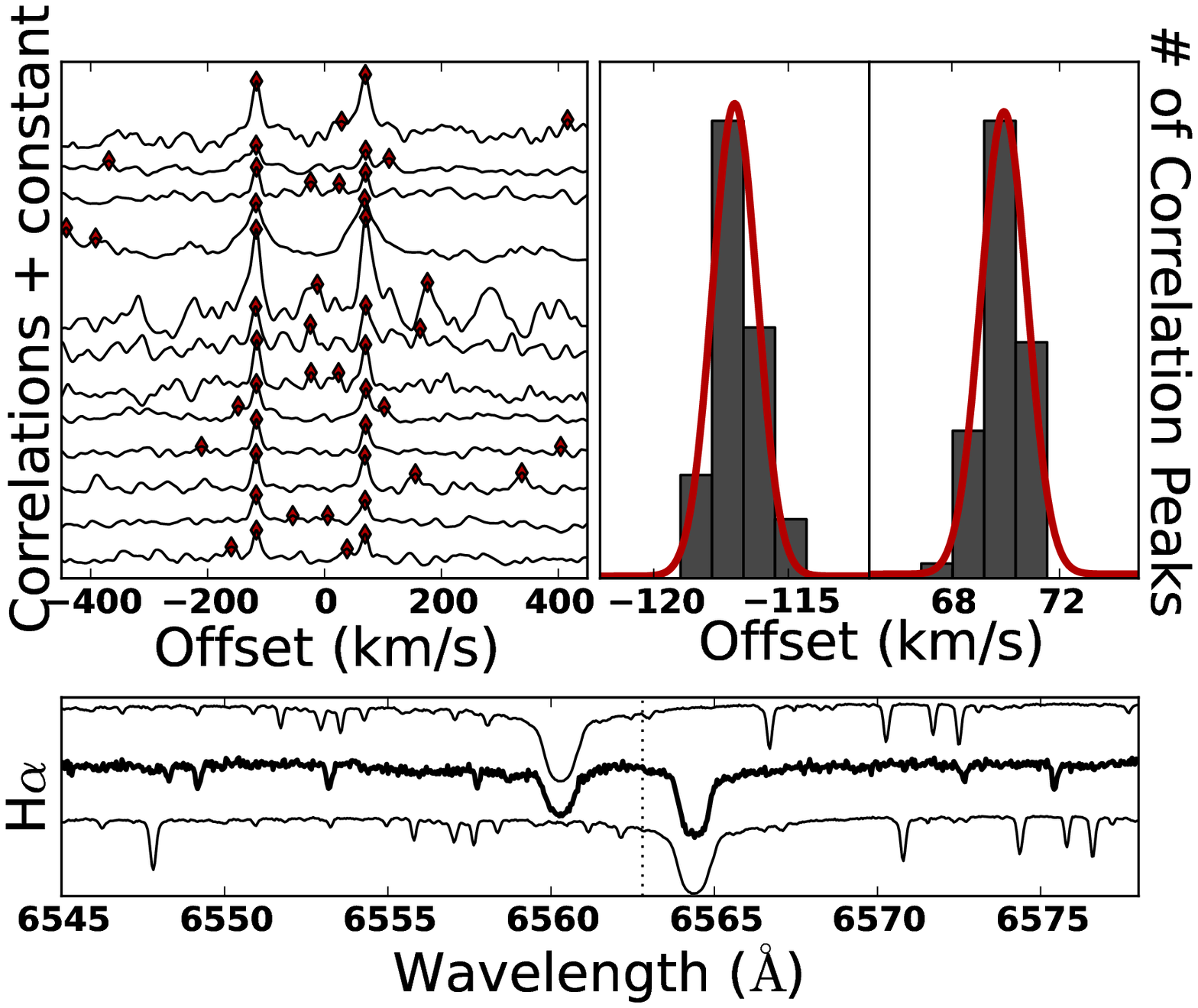}
\caption{Three steps in our cross-correlation pipeline, using a relatively high signal-to-noise
               observation of ASAS 205642+1153.0 as an example.
               The upper left panel shows the cross-correlation functions for several spectral chunks with 
               the four most significant peaks marked.
               The upper right panel shows zoomed in views of the histogram of significant peak values after cross-correlation of 
               all orders, as well as the best-fit Gaussians used to determine the velocities and errors.
               The lower panel shows the H$\alpha$ region of ASAS 205642+1153.0 (middle spectrum)
               along with that of the RV standard offset to match the best-fit RV values for each component.
               The dotted vertical line marks the rest wavelength of H$\alpha$.\label{fig:xcor}}
\end{figure}

After resampling the spectral chunks into bins equally spaced in $\ln \lambda$ (and oversampled by a factor of four), we applied a Bell-curve apodization to the ends of each chunk and cross-correlated each science spectrum with that of the RV standard source.  We record and histogram the RV values associated with the four most significant peaks of the correlation function of each spectral chunk and then fit a Gaussian to each major histogram peak to measure the RV and statistical errors (see Figure \ref{fig:xcor}).   Complete RV curves are reported in Table~\ref{tab:RV}.
Unfortunately the signal-to-noise ratio for our spectra, while sufficient for cross-correlation velocity measurements, are generally too low to provide meaningful measures of the rotational velocities of the component stars ($v$ sin $i$).  Note, as well, that a few of our spectra were taken near eclipse and those spectra may show blended line profiles (depending on the magnitude of the component $v$ sin $i$ values).  For a spectrum with very well-blended lines near eclipse, our pipeline would identify the approximate average RV of the two components as the primary (weighted by the relative line strengths and $v$ sin $i$ values of the components) and may additionally identify a spurious noise spike as the secondary star's RV; this is likely the case for one epoch of ASAS 073611-3123.4 observations (see Figure~\ref{fig:phoebe} and Table~\ref{tab:RV}).  The errors in our RV measurements are dominated by noise-driven variations in the correlation peak location, though line blending is likely also contributing to our measured errors for the few RV values measured around eclipse.

\subsection{Stellar parameters}
\label{sec:modeling}

\begin{figure*}
\centering
\includegraphics[width=\linewidth]{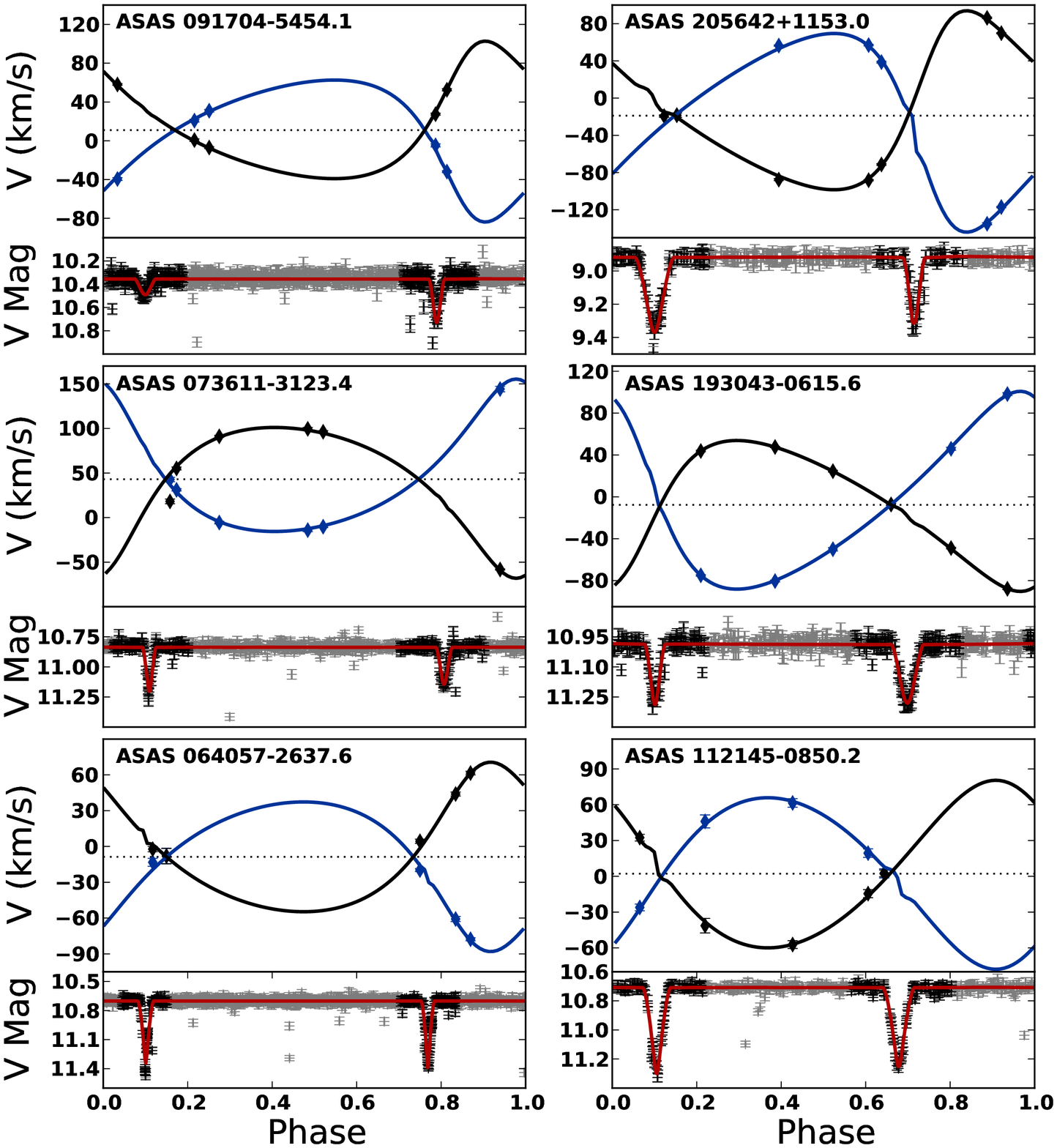}
\caption{Best-fit PHOEBE models for 6 high-$e$ DEBs. Best-fit model parameters for each system 
               are presented in Table~\ref{tab:params}.  
               For each RV curve the primary star is shown in black
               and the secondary in blue, with measurement errors included for each data point (oftentimes
               smaller than the plotted point).
               The dotted horizontal line marks the best-fit systemic velocity. The light curves
               include errors as reported by ASAS.  Points used in our PHOEBE fit are in black, points
               we excluded to speed up the fitting procedure are shown in grey, and the best-fit model is overlain in red.
 	     Note that we discarded a single RV observation of ASAS 073611-3123.4 at $\phi \approx 0.8$.
              The exact cause of this spurious point is unclear, but spectra observed near eclipse are plagued
               by spectral line blending; a likely culprit.
               \label{fig:phoebe}}
\end{figure*}

We used PHOEBE to fit physical models to the RV curves and light curves of 6 objects.  PHOEBE is a complete binary star modeling package based upon the Wilson-Devinney code 
\citep[WD;][]{WD1971, Wilson1979}.  We used version 0.32 of PHOEBE, which includes all of the features of the WD code as well as a graphical user interface and a 
few computational extensions \citep{phoebe}.
For each of the modeled systems we fit a PHOEBE detached binary model across 12 parameters until convergence was reached.  The 12 parameters we used were: 
semi-major orbital axis ($a$), mass ratio ($q$), systemic velocity ($v_{\gamma}$), orbital inclination ($i$), orbital eccentricity ($e$), orbital argument of periastron ($\omega$), 
the temperatures of both stars ($T_n$), the surface potentials of both stars ($\Omega_n$), the system's total luminosity, and the Modified Julian Date of a single primary eclipse ($t_{min,1}$).  For each system we adopt the period determined from the light curve (see \S\ref{subsec:pfind}), and we find very good agreement between the phase-folded light curves and RV curves despite the long delay between observations (light curves and RV curves were observed $\sim$10\,yr apart).
Note that we assumed any apsidal motion in these systems was negligible over the course of our observations, and we present the epoch of primary eclipse nearest the mean observation date of the photometric data.  To speed up the fitting procedure, we considered only the photometric data near eclipse (our lightcurves are very well sampled, and the out-of-eclipse points do little to constrain the model further).

The $\chi^2$ minimization procedure used by PHOEBE offers a straightforward way to estimate the formal errors on fitted parameters.  Assuming errors of the measured data are Gaussian, the inverse Hessian matrix of the fitted parameters approximates the covariance matrix of those parameters if the model is at the $\chi^2$ minimum.  After converging PHOEBE to a best-fit model, we follow the procedure outlined by the 
PHOEBE manual\footnote{\href{http://www.phoebe-project.org/?q=node/9}{PHOEBE V0.3x User Manual}},
and use the variances (estimated from the inverse Hessian by PHOEBE) to estimate our errors on all fitted and derived parameters.  

First, we reach convergence in PHOEBE, fitting all parameters to all of the data.  Then we estimate the formal errors for $a$, $q$ and $v_{\gamma}$ by calculating the Hessian matrix for a fit of $a$, $q$, $v_{\gamma}$, $t_{\min,1}$, $e$ and $\omega$ to only the RV curves.  This properly accounts for the most significant covariances amongst these variables and, because only the RV curves were used for this step, our formal errors are likely to be slightly overestimated.  Again starting from the converged model, we calculate the formal errors for all other fitted parameters through the Hessian matrix of a fit to all data while keeping $a$, $q$ and $v_{\gamma}$ fixed.

To estimate the formal errors on the derived masses and radii of these stars we do not attempt to propagate the formal errors of the fitted parameters, but instead simplify our task and differentiate the logarithm of the relevant equation, approximating $\sigma_p \approx |dp|$ for all parameters $p$.  Of course, we use Kepler's third law to estimate the errors on the masses and we use the expression of the generalized surface potential from \citet{WD1971} to estimate the errors of the radii.  (See the aforementioned PHOEBE manual for more details.) Our results are presented in Figure \ref{fig:phoebe} and best-fit parameters and errors for each system are listed in Table~\ref{tab:params}.

\section{Conclusion}
\label{sec:conclusion}

\begin{figure}
\centering
\includegraphics[width=\linewidth]{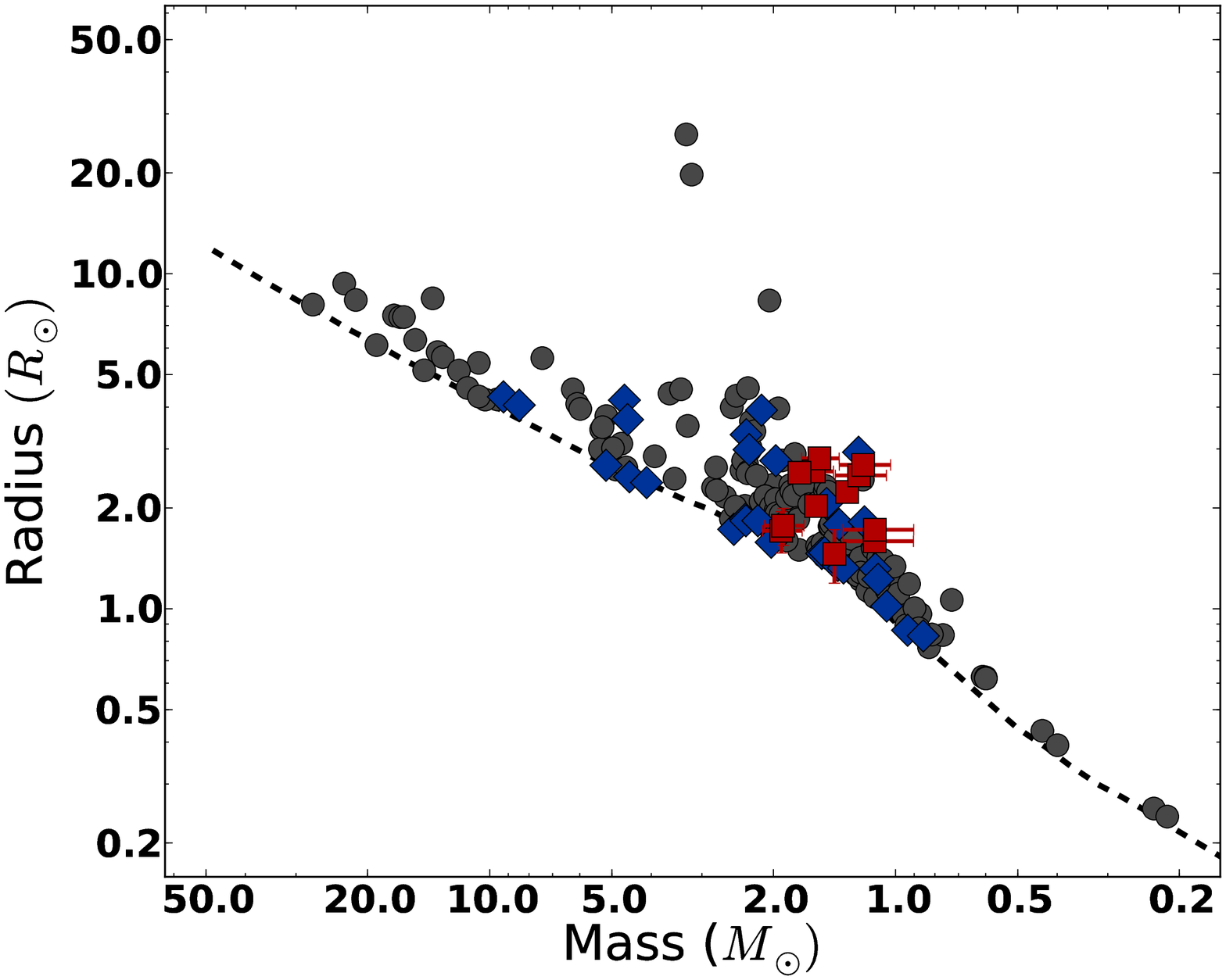}
\caption{Radii and masses for the 12 stars presented in Table~\ref{tab:everything} 
               and \citet{Torres}.
               Grey (circle): $e<0.1$ systems from \citet{Torres}; blue (diamond): $e>0.1$ systems from \citet{Torres};
               red (square): new systems presented in this work.  Note that errorbars are often smaller than the plotted symbol.
               The dashed line shows a theoretical zero-age main sequence for stars of solar metallicity \citep{girardi}.
               The 12 stars added here to this plot appear to fall within the normal ranges, though a complete
                analysis is beyond the scope of this paper.\label{fig:RvM}}
\end{figure}

Figure \ref{fig:eccents} displays the best-fit eccentricities for these sources alongside the results of previous studies of other high-$e$ binary systems.  
None of the systems presented here appear to be extreme outliers, though several do exhibit a remarkably high $e$ for their period.
Figure \ref{fig:RvM} plots the radii and masses of the 12 stars presented in Table~\ref{tab:params} 
alongside the results of previous studies.

Note that the population of stars included on that plot is heavily affected by several observational biases and any population studies should be
undertaken with caution.  
For example, only systems with relatively strong secondary eclipses are identified as DEBs by surveys like this one.  For this situation to arise a system with two main-sequence stars must have similar effective temperatures and therefore similar masses, which likely explains the nearly twin properties exhibited by some of the systems presented in this paper.  In contrast, binary systems with unequal masses will preferentially be detected only after the more massive primary star has moved away from the main sequence and expanded enough to reach an effective temperature comparable to that of the secondary star; this may help to explain the relatively large radii exhibited by a few of these stars presented here.
A complete analysis is beyond the scope of this paper; we encourage further studies to constrain the effect of tidal cicularization
on stellar binary components, and look forward to the important results coming from modern photometric surveys like Kepler 
\citep[e.g.][]{deboss13,frandsen13,hambleton} and the Palomar Transient Factory \citep[e.g.][]{Eyken11,Hamren11,Prince13}, as well as the next 
generation of synoptic photometric surveys.

We have outlined the process and results of a targeted search for highly-eccentric DEBs using the MACC probabilistic catalog and we presented 106 such systems with eccentricity estimates, most of which are listed here for the first time. We presented new RV curves and modeled masses and radii for 6 of these systems, each
of them double-lined spectroscopic binaries.  We publish these data to facilitate further studies of orbital evolution through tidal dissipation.  We also present this project as an example of how modern large-scale datasets can be immediately used to address outstanding problems in a wide range of astronomical subfields.  With limited funds and resources for new data acquisition, taking full advantage of the data already in hand and identifying the sources most worthy of further study is vitally important.  As we assemble these large-scale photometric datasets, we must also become more adept at understanding the data and locating rare objects of interest buried in them.

\FloatBarrier

\section*{Acknowledgments}
Thanks to Howard Isaacson, Erik Petigura and Lauren Weiss for helpful discussions about the Hamilton Spectrometer and spectral cross-correlation, to Geoff Marcy, Kelsey Clubb and the CHIRON team for their time and spectroscopic expertise, to Andrej Prsa for his expertise with PHOEBE, and to Adam Miller, Chris Klein and Dan Starr for helpful discussions about many things.  Thanks, as well, to our referee John Southworth for insightful comments that significantly improved this text.  The authors acknowledge the generous support of a CDI grant (\#0941742) from the National Science Foundation (NSF), as well as support from NSF grant AST-1211916 and the Richard and Rhoda Goldman Fund. This work includes data from the Cerro Tololo Inter-American Observatory, part of the National Optical Astronomy Observatory, which is operated by the Association of Universities for Research in Astronomy under contract with the National Science Foundation.

\bibliographystyle{apj}
\bibliography{bib}

\onecolumn 

\begin{deluxetable}{l l c c c}
\tablecolumns{5} 
\tabletypesize{\scriptsize}
\tablecaption{Journal of Spectroscopic Observations\label{tab:spectra}}
\tablewidth{0pt}
\tablehead{
\colhead{\bf{UT}} & \colhead{\bf{Object}} & \colhead{\bf{Instrument}} &
\colhead{\bf{Wavelegth Range (\AA)}} & \colhead{\bf{Exposure Time (s)}}
}
\startdata
2012-05-30 09:58:45.22 & ASAS-205642+1153.0 & HES & 3800--9600 & 600 \\ 
2012-05-30 11:49:37.30 & HD182488 & HES & 3800--9600 & 300 \\ 
2012-05-31 08:37:40.47 & ASAS-193043-0615.6 & HES & 3800--9600 & 900 \\ 
2012-05-31 08:50:13.87 & ASAS-205642+1153.0 & HES & 3800--9600 & 600 \\ 
2012-05-31 11:42:01.37 & HD182488 & HES & 3800--9600 & 300 \\ 
2012-06-01 08:26:27.41 & ASAS-193043-0615.6 & HES & 3800--9600 & 900 \\ 
2012-06-01 08:33:53.25 & ASAS-205642+1153.0 & HES & 3800--9600 & 300 \\ 
2012-06-01 11:47:44.47 & HD182488 & HES & 3800--9600 & 2 \\ 
2012-06-02 08:34:07.58 & ASAS-193043-0615.6 & HES & 3800--9600 & 600 \\ 
2012-06-02 08:42:52.43 & ASAS-205642+1153.0 & HES & 3800--9600 & 300 \\ 
2012-06-02 11:49:31.78 & HD182488 & HES & 3800--9600 & 2 \\ 
2012-06-03 09:02:30.76 & ASAS-193043-0615.6 & HES & 3800--9600 & 600 \\ 
2012-06-03 09:08:12.68 & ASAS-205642+1153.0 & HES & 3800--9600 & 180 \\ 
2012-06-03 11:46:03.57 & HD182488 & HES & 3800--9600 & 10 \\ 
2012-06-04 08:47:39.36 & ASAS-193043-0615.6 & HES & 3800--9600 & 600 \\ 
2012-06-04 08:53:25.38 & ASAS-205642+1153.0 & HES & 3800--9600 & 180 \\ 
2012-06-06 08:27:51.06 & ASAS-193043-0615.6 & HES & 3800--9600 & 900 \\ 
2012-06-06 08:40:30.66 & ASAS-205642+1153.0 & HES & 3800--9600 & 600 \\ 
2012-06-06 11:40:09.16 & HD182488 & HES & 3800--9600 & 300 \\ 
2012-11-04 07:14:09.4 & ASAS-064057-2637.6 & CHIRON & 4600--8800 & 500 \\ 
2012-11-08 05:43:08.7 & ASAS-064057-2637.6 & CHIRON & 4600--8800 & 500 \\ 
2012-11-10 09:04:53.5 & ASAS-073611-3123.4 & CHIRON & 4600--8800 & 500 \\ 
2012-11-13 06:53:00.5 & ASAS-073611-3123.4 & CHIRON & 4600--8800 & 500 \\ 
2012-11-17 04:51:32.6 & ASAS-064057-2637.6 & CHIRON & 4600--8800 & 500 \\ 
2012-11-17 07:18:49.3 & ASAS-073611-3123.4 & CHIRON & 4600--8800 & 500 \\ 
2012-11-20 06:40:54.2 & ASAS-064057-2637.6 & CHIRON & 4600--8800 & 500 \\ 
2012-11-20 06:51:20.2 & ASAS-073611-3123.4 & CHIRON & 4600--8800 & 500 \\ 
2012-11-20 07:02:10.0 & HD75289 & CHIRON & 4600--8800 & 60 \\ 
2012-11-28 03:40:30.1 & ASAS-064057-2637.6 & CHIRON & 4600--8800 & 500 \\ 
2012-11-28 05:52:42.7 & ASAS-073611-3123.4 & CHIRON & 4600--8800 & 500 \\ 
2012-11-28 06:14:28.8 & ASAS-091704-5454.1 & CHIRON & 4600--8800 & 500 \\ 
2012-11-29 07:59:39.3 & ASAS-112145-0850.2 & CHIRON & 4600--8800 & 500 \\ 
2012-12-01 07:19:43.1 & ASAS-073611-3123.4 & CHIRON & 4600--8800 & 500 \\ 
2012-12-01 07:41:43.2 & ASAS-091704-5454.1 & CHIRON & 4600--8800 & 500 \\ 
2012-12-02 08:12:48.1 & ASAS-112145-0850.2 & CHIRON & 4600--8800 & 500 \\ 
2012-12-10 07:03:51.7 & ASAS-064057-2637.6 & CHIRON & 4600--8800 & 500 \\ 
2012-12-16 08:30:01.2 & ASAS-091704-5454.1 & CHIRON & 4600--8800 & 500 \\ 
2012-12-17 08:23:54.3 & ASAS-112145-0850.2 & CHIRON & 4600--8800 & 500 \\ 
2012-12-23 07:16:27.1 & ASAS-091704-5454.1 & CHIRON & 4600--8800 & 500 \\ 
2012-12-28 06:54:01.1 & ASAS-091704-5454.1 & CHIRON & 4600--8800 & 500 \\ 
2012-12-29 06:27:20.9 & ASAS-112145-0850.2 & CHIRON & 4600--8800 & 500 \\ 
2013-01-06 07:54:27.3 & ASAS-112145-0850.2 & CHIRON & 4600--8800 & 500 \\ 
\enddata
\tablecomments{A log of the spectroscopic observations obtained for this project.  
                              Data were collected on the SMARTS 1.5\,m telescope at CTIO (CHIRON)
                              or on the 3\,m Shane telescope at Mt.\ Hamilton (HES). 
               Note that observations of the RV standard HD182488 were taken with widely varying
                exposure times on different nights. Different exposures exhibit different signal levels, but all are high enough quality to perform
                cross-correlation and none are saturated. For all HES spectra we chose to use the
                standard observations taken the same night as science observations.}
\end{deluxetable}
\clearpage

\begin{deluxetable}{c c c r c | c c c r c}
\tablecolumns{10} 
\tabletypesize{\scriptsize}
\tablecaption{106 Eccentric Eclipsing Binaries in ASAS/MACC\label{tab:everything}}
\tablewidth{0pt}
\tablehead{
\colhead{ASAS ID} & \colhead{$\alpha$ (J2000) } & \colhead{$\delta$ (J2000) } &
\colhead{ P (d) } & \colhead{e} & 
\colhead{ASAS ID} & \colhead{$\alpha$ (J2000) } & \colhead{$\delta$ (J2000) } &
\colhead{ P (d) } & \colhead{e}
}
\startdata
144242-5904.1 & 14:42:39.758 & $-$59:04:17.80 & 7.2725 & $0.64\pm0.017$ & 122729-6411.9 & 12:27:28.843 & $-$64:12:05.40 & 15.8034 & $0.52\pm0.109$\\ 
074935-2415.8 & 07:49:35.198 & $-$24:15:50.98 & 7.2738 & $0.50\pm0.033$ & 153921-6053.6 & 15:39:22.540 & $-$60:53:22.09 & 8.6013 & $0.50\pm0.016$\\ 
185105-1818.7 & 18:51:05.393 & $-$18:18:45.42 & 4.5623 & $0.48\pm0.038$ & 142933-6014.6 & 14:29:32.543 & $-$60:14:36.09 & 14.1678 & $0.46\pm0.022$\\ 
100433-6315.7 & 10:04:32.855 & $-$63:15:43.23 & 13.8079 & $0.46\pm0.013$ & 151543-2035.5 & 15:15:43.312 & $-$20:35:26.69 & 17.2088 & $0.44\pm0.041$\\ 
073032-1611.7 & 07:30:32.231 & $-$16:11:40.11 & 2.6862 & $0.43\pm0.139$ & 061727+1100.3 & 06:17:27.355 & $+$11:00:17.60 & 15.7199 & $0.41\pm0.030$\\ 
072903-1525.4 & 07:29:03.426 & $-$15:25:22.74 & 8.8512 & $0.40\pm0.070$ & 132002-6144.9 & 13:20:01.957 & $-$61:44:51.54 & 7.3086 & $0.40\pm0.041$\\ 
072515-1135.9 & 07:25:15.017 & $-$11:35:49.62 & 4.5100 & $0.39\pm0.022$ & 122540-6337.6 & 12:25:40.091 & $-$63:37:33.48 & 3.0816 & $0.39\pm0.097$\\ 
100625-5500.8 & 10:06:25.333 & $-$55:00:44.63 & 4.4112 & $0.39\pm0.063$ & 175630-5509.7 & 17:56:29.566 & $-$55:09:41.50 & 11.1781 & $0.38\pm0.083$\\ 
081145-2519.8 & 08:11:44.808 & $-$25:19:46.56 & 9.7995 & $0.38\pm0.017$ & 084424-3715.4 & 08:44:23.784 & $-$37:15:24.78 & 5.7289 & $0.37\pm0.077$\\ 
091704-5454.1$^{\dag}$ & 09:17:03.746 & $-$54:54:04.52 & 12.3547 & $0.37\pm0.004$ & 134702-6237.1 & 13:47:03.091 & $-$62:36:57.37 & 3.8729 & $0.36\pm0.149$\\ 
102732-6400.4 & 10:27:31.910 & $-$64:00:21.78 & 5.6469 & $0.36\pm0.145$ & 083612-4547.7 & 08:36:11.340 & $-$45:47:38.61 & 6.9552 & $0.36\pm0.058$\\ 
151308-2504.3 & 15:13:07.993 & $-$25:04:18.18 & 12.7256 & $0.35\pm0.094$ & 180743-2712.0 & 18:07:43.558 & $-$27:12:02.09 & 6.9237 & $0.35\pm0.028$\\ 
075942-1411.2 & 07:59:41.500 & $-$14:11:11.15 & 4.4058 & $0.35\pm0.241$ & 070127-0307.0 & 07:01:27.188 & $-$03:07:02.40 & 21.9670 & $0.34\pm0.028$\\ 
205642+1153.0$^{\dag}$ & 20:56:41.986 & $+$11:53:02.37 & 4.1244 & $0.34\pm0.003$ & 083400-3431.4 & 08:33:59.630 & $-$34:31:35.20 & 3.1376 & $0.33\pm0.051$\\ 
073611-3123.4$^{\dag}$ & 07:36:10.706 & $-$31:23:21.72 & 8.8550 & $0.33\pm0.002$ & 081731-6043.8 & 08:17:30.077 & $-$60:43:35.25 & 8.3773 & $0.33\pm0.041$\\ 
074401-3105.4 & 07:44:00.456 & $-$31:05:24.64 & 6.2042 & $0.32\pm0.015$ & 062926-2513.5 & 06:29:25.307 & $-$25:13:40.10 & 26.3808 & $0.32\pm0.073$\\ 
094810-5117.4 & 09:48:10.152 & $-$51:17:21.87 & 7.1491 & $0.32\pm0.060$ & 164753-3239.7 & 16:47:54.283 & $-$32:39:30.69 & 3.9443 & $0.31\pm0.080$\\ 
193043-0615.6$^{\dag}$ & 19:30:43.031 & $-$06:15:34.80 & 7.2764 & $0.31\pm0.003$ & 111013-6051.8 & 11:10:12.378 & $-$60:51:47.17 & 5.3829 & $0.31\pm0.074$\\ 
062005+0454.8 & 06:20:03.728 & $+$04:54:46.77 & 5.9415 & $0.30\pm0.043$ & 144442-7721.9 & 14:44:41.078 & $-$77:21:53.00 & 10.0220 & $0.30\pm0.024$\\ 
175333-2031.2 & 17:53:32.946 & $-$20:31:09.49 & 3.3157 & $0.30\pm0.039$ & 063002-4959.3 & 06:30:02.704 & $-$49:59:17.11 & 41.7455 & $0.30\pm0.008$\\ 
140737-6957.6 & 14:07:37.196 & $-$69:57:34.13 & 6.5166 & $0.29\pm0.128$ & 072222-1159.8 & 07:22:21.691 & $-$11:59:45.77 & 13.0144 & $0.29\pm0.025$\\ 
091313-4744.4 & 09:13:12.907 & $-$47:44:24.00 & 2.4160 & $0.29\pm0.132$ & 083150-1529.0 & 08:31:49.829 & $-$15:29:00.06 & 8.7409 & $0.29\pm0.030$\\ 
180814-1712.1 & 18:08:14.579 & $-$17:12:02.53 & 3.3342 & $0.29\pm0.079$ & 171145-2819.0 & 17:11:45.035 & $-$28:19:01.70 & 3.1451 & $0.29\pm0.027$\\ 
122743-6224.0 & 12:27:45.785 & $-$62:23:46.67 & 15.5501 & $0.28\pm0.023$ & 100243-5643.4 & 10:02:43.760 & $-$56:43:07.71 & 10.9401 & $0.28\pm0.068$\\ 
162754-5741.6 & 16:27:52.114 & $-$57:41:41.85 & 3.2236 & $0.28\pm0.032$ & 193934-1740.0 & 19:39:35.086 & $-$17:39:39.22 & 10.7034 & $0.28\pm0.055$\\ 
110812-3014.6 & 11:08:11.634 & $-$30:14:40.19 & 6.2149 & $0.28\pm0.110$ & 133533-5214.5 & 13:35:32.932 & $-$52:14:30.76 & 11.4405 & $0.28\pm0.008$\\ 
064057-2637.6$^{\dag}$ & 06:40:57.079 & $-$26:37:34.55 & 12.4460 & $0.28\pm0.001$ & 082428-4525.2 & 08:24:28.026 & $-$45:25:11.52 & 3.4511 & $0.28\pm0.070$\\ 
183356-2922.7 & 18:33:55.645 & $-$29:22:40.60 & 14.8986 & $0.28\pm0.118$ & 183020-0923.2 & 18:30:20.581 & $-$09:23:09.54 & 20.1221 & $0.26\pm0.062$\\ 
122731-5549.4 & 12:27:31.284 & $-$55:49:22.39 & 9.8812 & $0.26\pm0.015$ & 104739-6037.1 & 10:47:37.352 & $-$60:36:53.86 & 17.7990 & $0.25\pm0.023$\\ 
173627-3329.6 & 17:36:27.400 & $-$33:29:35.86 & 23.4448 & $0.25\pm0.042$ & 183235-3314.7 & 18:32:34.130 & $-$33:14:54.18 & 3.6899 & $0.25\pm0.124$\\ 
083903-4343.9 & 08:39:03.316 & $-$43:43:55.81 & 5.3297 & $0.24\pm0.042$ & 062114-0534.8 & 06:21:14.429 & $-$05:34:48.18 & 2.4638 & $0.24\pm0.082$\\ 
060719+1331.8 & 06:07:18.599 & $+$13:31:45.80 & 5.0780 & $0.23\pm0.236$ & 112926-6202.0 & 11:29:26.430 & $-$62:01:57.94 & 3.2249 & $0.23\pm0.353$\\ 
093052-3234.3 & 09:30:51.541 & $-$32:34:21.59 & 4.0217 & $0.22\pm0.058$ & 152227-4718.6 & 15:22:26.771 & $-$47:18:34.17 & 24.2154 & $0.22\pm0.022$\\ 
085427-4135.0 & 08:54:26.968 & $-$41:34:58.14 & 3.4523 & $0.22\pm0.069$ & 165354-1301.9 & 16:53:54.193 & $-$13:01:57.53 & 2.2076 & $0.22\pm0.051$\\ 
081855-3612.8 & 08:18:55.325 & $-$36:12:48.33 & 9.3535 & $0.21\pm0.089$ & 173319-4315.1 & 17:33:21.110 & $-$43:15:17.84 & 12.7766 & $0.21\pm0.026$\\ 
050351-1541.9 & 05:03:51.145 & $-$15:41:53.91 & 7.2792 & $0.20\pm0.093$ & 111507-4815.6 & 11:15:06.980 & $-$48:15:33.37 & 6.7868 & $0.20\pm0.102$\\ 
075207-3032.2 & 07:52:08.371 & $-$30:32:18.69 & 10.6929 & $0.20\pm0.044$ & 180908-2128.5 & 18:09:06.556 & $-$21:28:26.71 & 2.2703 & $0.20\pm0.051$\\ 
135817-3004.5 & 13:58:17.270 & $-$30:04:35.60 & 10.2863 & $0.20\pm0.064$ & 065554-0014.4 & 06:55:53.555 & $-$00:14:23.70 & 2.9707 & $0.20\pm0.036$\\ 
075052+0048.0 & 07:50:51.338 & $+$00:48:03.94 & 2.0543 & $0.20\pm0.079$ & 151510-5928.8 & 15:15:09.864 & $-$59:28:47.68 & 8.7061 & $0.19\pm0.139$\\ 
103449-6013.1 & 10:34:49.148 & $-$60:13:04.27 & 7.5939 & $0.19\pm0.030$ & 054218+2003.7 & 05:42:18.029 & $+$20:03:39.18 & 5.1157 & $0.19\pm0.066$\\ 
105835+0329.4 & 10:58:34.777 & $+$03:29:20.20 & 24.3757 & $0.18\pm0.069$ & 073218-1436.5 & 07:32:17.873 & $-$14:36:29.19 & 8.1780 & $0.18\pm0.080$\\ 
181641-3538.2 & 18:16:41.585 & $-$35:38:09.70 & 4.3028 & $0.18\pm0.041$ & 100104-5847.5 & 10:01:04.084 & $-$58:47:25.94 & 14.3202 & $0.18\pm0.038$\\ 
194116+1301.3 & 19:41:16.098 & $+$13:01:18.72 & 1.5609 & $0.17\pm0.063$ & 185951-4711.8 & 18:59:51.288 & $-$47:11:47.64 & 2.8777 & $0.17\pm0.065$\\ 
063202-6750.7 & 06:32:01.802 & $-$67:50:38.18 & 5.8191 & $0.17\pm0.098$ & 091227-3801.6 & 09:12:27.443 & $-$38:01:39.61 & 4.6252 & $0.17\pm0.033$\\ 
081237-1356.9 & 08:12:36.439 & $-$13:57:00.62 & 9.6469 & $0.15\pm0.061$ & 091038-4304.9 & 09:10:37.733 & $-$43:04:56.04 & 4.6701 & $0.15\pm0.014$\\ 
150046-5014.9 & 15:00:47.614 & $-$50:15:07.92 & 9.2879 & $0.15\pm0.186$ & 100421-3319.0 & 10:04:20.626 & $-$33:19:00.74 & 6.7148 & $0.15\pm0.033$\\ 
064753-1642.9 & 06:47:52.566 & $-$16:42:56.34 & 1.8434 & $0.13\pm0.066$ & 084503-4113.0 & 08:45:01.753 & $-$41:12:59.03 & 2.0666 & $0.13\pm0.023$\\ 
112145-0850.2$^{\dag}$ & 11:21:44.284 & $-$08:50:00.21 & 8.3687 & $0.13\pm0.003$ & 151146-5416.4 & 15:11:45.956 & $-$54:16:26.53 & 2.7021 & $0.13\pm0.063$\\ 
111440-6039.6 & 11:14:42.302 & $-$60:39:25.97 & 2.4222 & $0.13\pm0.056$ & 070911+1211.2 & 07:09:10.840 & $+$12:11:19.05 & 2.2777 & $0.13\pm0.078$\\ 
075509-2310.4 & 07:55:09.102 & $-$23:10:23.76 & 3.9511 & $0.13\pm0.149$ & 115007-6425.1 & 11:50:06.990 & $-$64:25:05.10 & 6.9050 & $0.12\pm0.102$\\ 
165336-4615.9 & 16:53:36.377 & $-$46:15:53.64 & 2.3302 & $0.12\pm0.045$ & 142655-6735.3 & 14:26:55.057 & $-$67:35:18.05 & 3.8769 & $0.12\pm0.079$\\ 
172034-4206.2 & 17:20:33.619 & $-$42:06:11.16 & 5.5398 & $0.11\pm0.172$ & 061016-3321.3 & 06:10:15.636 & $-$33:21:36.12 & 200.0717 & $0.11\pm0.039$\\ 
164208-3346.0 & 16:42:08.093 & $-$33:46:00.56 & 2.4916 & $0.11\pm0.021$ & 074123+0253.3 & 07:41:22.376 & $+$02:53:05.37 & 8.5995 & $0.10\pm0.060$
\enddata
\tablecomments{The 106 high-$e$ systems identified in this paper, with accurate periods and eccentricity estimates as determined by light curve
                             fits with the DEBiL code. Characteristic errors on the periods are about $\pm 0.006$\,d \\
                             $^{\dag}$Objects with new RV curves and detailed models presented in \S \ref{sec:RV}.}
\end{deluxetable}
\clearpage

\begin{deluxetable}{r c c r c c c}
\tablecolumns{7} 
\tabletypesize{\scriptsize}
\tablecaption{New Radial Velocities for 6 Systems\label{tab:RV}}
\tablewidth{0pt}
\tablehead{
\colhead{UT} & \colhead{V$_1 \pm 1\sigma$} & \colhead{V$_2 \pm 1\sigma$} &
\colhead{Standard} & \colhead{V$_{std} \pm 1 \sigma$} &
\colhead{BCC$_{star}$} & \colhead{BCC$_{std}$}
}
\startdata
\multicolumn{7}{ c }{\bf{ASAS 091704-5454.1 (HD 299972)}} \\ 
\tableline
2012-11-28 06:14:28.8 & 27.4$\pm$1.07 & -4.8$\pm$1.64 & HD75289 & 9.9$\pm$0.64$^*$ & 12.31 & 16.67 \\
2012-12-01 07:41:43.2 & 58.1$\pm$0.88 & -39.5$\pm$1.47 & HD75289 & 9.9$\pm$0.64 & 12.53 & 16.67 \\
2012-12-16 08:30:01.2 & -7.0$\pm$0.96 & 31.0$\pm$1.34 & HD75289 & 9.9$\pm$0.64 & 13.18 & 16.67 \\
2012-12-23 07:16:27.1 & 52.6$\pm$1.10 & -31.8$\pm$0.94 & HD75289 & 9.9$\pm$0.64 & 13.16 & 16.67 \\
2012-12-28 06:54:01.1 & 0.7$\pm$1.32 & 20.8$\pm$1.47 & HD75289 & 9.9$\pm$0.64 & 13.02 & 16.67 \\
\tableline
\multicolumn{7}{ c }{\bf{ASAS 205642+1153.0 (HD 199428)}} \\ 
\tableline
2012-05-30 09:58:45.22 & 69.8$\pm$0.56 & -117.1$\pm$0.65 & HD182488 & -21.6$\pm$0.07$^{\dag}$ & 24.67 & 13.49 \\
2012-05-31 08:50:13.87 & -18.4$\pm$2.33 & --- & HD182488 & -21.6$\pm$0.07 & 24.53 & 13.31 \\
2012-06-01 08:33:53.25 & -87.6$\pm$1.12 & 56.4$\pm$0.70 & HD182488 & -21.6$\pm$0.07 & 24.38 & 13.12 \\
2012-06-02 08:42:52.43 & -71.3$\pm$0.88 & 38.7$\pm$0.97 & HD182488 & -21.6$\pm$0.07 & 24.22 & 12.93 \\
2012-06-03 09:08:12.68 & 86.2$\pm$0.79 & -135.1$\pm$1.10 & HD182488 & -21.6$\pm$0.07 & 24.05 & 12.74 \\
2012-06-04 08:53:25.38 & -19.3$\pm$1.32 & --- & HD182488 & -21.6$\pm$0.07 & 23.89 & 12.74 \\
2012-06-06 08:40:30.66 & -88.2$\pm$0.74 & 57.0$\pm$0.78 & HD182488 & -21.6$\pm$0.07 & 23.53 & 12.15 \\
\tableline
\multicolumn{7}{ c }{\bf{ASAS 073611-3123.4 (TYC 7105-2473-1)}} \\ 
\tableline
2012-11-10 09:04:53.5 & 18.2$\pm$0.64$^{\ddag}$ & 42.7$\pm$3.06 & HD75289 & 9.9$\pm$0.64 & 17.93 & 16.67 \\
2012-11-13 06:53:00.5 & 99.5$\pm$1.68 & -14.1$\pm$1.69 & HD75289 & 9.9$\pm$0.64 & 17.69 & 16.67 \\
2012-11-17 07:18:49.3 & -58.2$\pm$1.06 & 144.2$\pm$1.38 & HD75289 & 9.9$\pm$0.64 & 17.26 & 16.67 \\
2012-11-20 06:51:20.2 & 91.1$\pm$1.60 & -5.6$\pm$1.32 & HD75289 & 9.9$\pm$0.64 & 16.89 & 16.67 \\
2012-11-28 05:52:42.7 & 55.1$\pm$2.63 & 31.1$\pm$2.42 & HD75289 & 9.9$\pm$0.64 & 15.69 & 16.67 \\
2012-12-01 07:19:43.1 & 96.2$\pm$1.23 & -10.2$\pm$0.97 & HD75289 & 9.9$\pm$0.64 & 15.17 & 16.67 \\
\tableline
\multicolumn{7}{ c }{\bf{ASAS 193043-0615.6 (TYC 5156-179-1)}} \\ 
\tableline
2012-05-31 08:37:40.47 & 47.6$\pm$0.57 & -80.5$\pm$0.86 & HD182488 & -21.6$\pm$0.07 & 19.63 & 13.31 \\
2012-06-01 08:26:27.41 & 24.5$\pm$0.73 & -50.1$\pm$1.54 & HD182488 & -21.6$\pm$0.07 & 19.28 & 13.12 \\
2012-06-02 08:34:07.58 & -7.3$\pm$0.75 & --- & HD182488 & -21.6$\pm$0.07 & 18.92 & 12.93 \\
2012-06-03 09:02:30.76 & -48.9$\pm$0.81 & 45.9$\pm$0.92 & HD182488 & -21.6$\pm$0.07 & 18.55 & 12.74 \\
2012-06-04 08:47:39.36 & -87.9$\pm$0.79 & 98.2$\pm$1.46 & HD182488 & -21.6$\pm$0.07 & 18.19 & 12.74 \\
2012-06-06 08:27:51.06 & 43.8$\pm$0.77 & -75.0$\pm$1.36 & HD182488 & -21.6$\pm$0.07 & 17.45 & 12.15 \\
\tableline
\multicolumn{7}{ c }{\bf{ASAS 064057-2637.6 (TYC 6516-1884-1)}} \\ 
\tableline
2012-11-04 07:14:09.4 & 44.0$\pm$1.68 & -60.8$\pm$2.11 & HD75289 & 9.9$\pm$0.64 & 17.08 & 16.67 \\
2012-11-08 05:43:08.7 & -8.0$\pm$6.49 & --- & HD75289 & 9.9$\pm$0.64 & 16.42 & 16.67 \\
2012-11-17 04:51:32.6 & 61.6$\pm$1.51 & -77.8$\pm$1.30 & HD75289 & 9.9$\pm$0.64 & 14.62 & 16.67 \\
2012-11-20 06:40:54.2 & -2.2$\pm$2.22 & -13.3$\pm$3.31 & HD75289 & 9.9$\pm$0.64 & 13.90 & 16.67 \\
2012-11-28 03:40:30.1 & 4.3$\pm$1.74 & -19.8$\pm$1.26 & HD75289 & 9.9$\pm$0.64 & 11.91 & 16.67 \\
\tableline
\multicolumn{7}{ c }{\bf{ASAS 112145-0850.2 (BD-08 3152)}} \\ 
\tableline
2012-11-29 07:59:39.3 & 32.4$\pm$6.07 & -26.1$\pm$5.45 & HD75289 & 9.9$\pm$0.64 & 28.25 & 16.67 \\
2012-12-02 08:12:48.1 & -57.2$\pm$3.40 & 61.5$\pm$3.57 & HD75289 & 9.9$\pm$0.64 & 28.66 & 16.67 \\
2012-12-17 08:23:54.3 & -41.4$\pm$3.29 & 46.0$\pm$3.58 & HD75289 & 9.9$\pm$0.64 & 29.61 & 16.67 \\
2012-12-29 06:27:20.9 & 2.5$\pm$2.81 & --- & HD75289 & 9.9$\pm$0.64 & 28.84 & 16.67 \\
2013-01-06 07:54:27.3 & -14.5$\pm$2.79 & 19.4$\pm$3.83 & HD75289 & 9.9$\pm$0.64 & 27.63 & 16.67 \\
\enddata
\tablecomments{Radial velocities for science and standard targets used in this project, as described in \S \ref{sec:RV}. 
               Epochs observed near eclipse (line-blended spectra) are listed with only a single RV measurement.\\
                              $^*$\citet{HD75289} \\
                              $^{\dag}$\citet{HD182488} \\
                              $^{\ddag}$Spurious point; disregarded in PHOEBE fit
}
\end{deluxetable}
\clearpage


\begin{table*}
\centering
\begingroup
\fontsize{8pt}{10pt}\selectfont
\caption{Modeled parameters for 6 systems\label{tab:params}}
\begin{multicols}{3}

	\begin{tabular}{l r}
	\hline
	\multicolumn{2}{ c }{\bf{ASAS 091704-5454.1}} \\
	\multicolumn{2}{ c }{ \bf{(HD 299972)}} \\
	\hline
	M$_1$ (M$_{\sun}$) & 1.59$\pm$0.17 \\
	M$_2$              & 1.54$\pm$0.16 \\
	R$_1$ (R$_{\sun}$) & 2.57$\pm$0.09 \\
	R$_2$              & 2.81$\pm$0.11 \\
	Period (d) & 12.3547$\pm$0.0111 \\
	t$_{min,1}$ (MJD) & 53380.643$\pm$0.009 \\
	a (R$_{\sun}$) & 32.9$\pm$0.9 \\
	i ($^{\circ}$) & 84.85$\pm$0.21 \\
	e              & 0.3658$\pm$0.0041 \\
	$\omega$ ($^{\circ}$) & 143.67$\pm$0.85 \\
	v$_{\gamma}$ (\kms) & 10.97$\pm$0.60 \\
	\end{tabular}

	\begin{tabular}{l r}
	\hline
	\multicolumn{2}{ c }{\bf{ASAS 205642+1153.0}} \\
	\multicolumn{2}{ c }{ \bf{(HD 199428)}} \\
	\hline
	M$_1$ (M$_{\sun}$) & 1.57$\pm$0.07 \\
	M$_2$              & 1.41$\pm$0.06 \\
	R$_1$ (R$_{\sun}$) & 2.03$\pm$0.09 \\
	R$_2$              & 1.46$\pm$0.27 \\
	Period (d) & 4.1244$\pm$0.0008 \\
	t$_{min,1}$ (MJD) & 53896.308$\pm$0.003 \\
	a (R$_{\sun}$) & 15.6$\pm$0.2 \\
	i ($^{\circ}$) & 87.49$\pm$0.80 \\
	e              & 0.3396$\pm$0.0031 \\
	$\omega$ ($^{\circ}$) & 120.54$\pm$0.40 \\
	v$_{\gamma}$ (\kms) & -18.88$\pm$0.42 \\
	\end{tabular}

	\begin{tabular}{l r}
	\hline
	\multicolumn{2}{ c }{\bf{ASAS 073611-3123.4}} \\
	\multicolumn{2}{ c }{ \bf{(TYC 7105-2473-1)}} \\
	\hline
	M$_1$ (M$_{\sun}$) & 1.91$\pm$0.21 \\
	M$_2$              & 1.89$\pm$0.21 \\
	R$_1$ (R$_{\sun}$) & 1.71$\pm$0.24 \\
	R$_2$              & 1.77$\pm$0.22 \\
	Period (d) & 8.8550$\pm$0.0088 \\
	t$_{min,1}$ (MJD) & 53478.189$\pm$0.005 \\
	a (R$_{\sun}$) & 28.1$\pm$1.0 \\
	i ($^{\circ}$) & 87.02$\pm$0.24 \\
	e              & 0.3323$\pm$0.0020 \\
	$\omega$ ($^{\circ}$) & 19.01$\pm$0.96 \\
	v$_{\gamma}$ (\kms) & 43.07$\pm$0.20 \\
	\end{tabular}

	\begin{tabular}{l r}
	\hline
	\multicolumn{2}{ c }{\bf{ASAS 193043-0615.6}} \\
	\multicolumn{2}{ c }{ \bf{(TYC 5156-179-1)}} \\
	\hline
	M$_1$ (M$_{\sun}$) & 1.73$\pm$0.02 \\
	M$_2$              & 1.32$\pm$0.02 \\
	R$_1$ (R$_{\sun}$) & 2.55$\pm$0.05 \\
	R$_2$              & 2.23$\pm$0.06 \\
	Period (d) & 7.2764$\pm$0.0016 \\
	t$_{min,1}$ (MJD) & 53536.800$\pm$0.002 \\
	a (R$_{\sun}$) & 22.9$\pm$0.1 \\
	i ($^{\circ}$) & 85.27$\pm$0.13 \\
	e              & 0.3072$\pm$0.0026 \\
	$\omega$ ($^{\circ}$) & 61.27$\pm$0.34 \\
	v$_{\gamma}$ (\kms) & -7.62$\pm$0.14 \\
	\end{tabular}

	\begin{tabular}{l r}
	\hline
	\multicolumn{2}{ c }{\bf{ASAS 064057-2637.6}} \\
	\multicolumn{2}{ c }{ \bf{(TYC 6516-1884-1)}} \\
	\hline
	M$_1$ (M$_{\sun}$) & 1.13$\pm$0.22 \\
	M$_2$              & 1.12$\pm$0.22 \\
	R$_1$ (R$_{\sun}$) & 1.59$\pm$0.12 \\
	R$_2$              & 1.72$\pm$0.13 \\
	Period (d) & 12.4460$\pm$0.0093 \\
	t$_{min,1}$ (MJD) & 53197.648$\pm$0.004 \\
	a (R$_{\sun}$) & 29.6$\pm$1.8 \\
	i ($^{\circ}$) & 90.40$\pm$0.29 \\
	e              & 0.2799$\pm$0.0013 \\
	$\omega$ ($^{\circ}$) & 161.90$\pm$0.58 \\
	v$_{\gamma}$ (\kms) & -8.70$\pm$0.90 \\
	\end{tabular}

	\begin{tabular}{l r}
	\hline
	\multicolumn{2}{ c }{\bf{ASAS 112145-0850.2}} \\
	\multicolumn{2}{ c }{ \bf{(BD-08 3152)}} \\
	\hline
	M$_1$ (M$_{\sun}$) & 1.23$\pm$0.18 \\
	M$_2$              & 1.20$\pm$0.17 \\
	R$_1$ (R$_{\sun}$) & 2.50$\pm$0.04 \\
	R$_2$              & 2.69$\pm$0.04 \\
	Period (d) & 8.3687$\pm$0.0019 \\
	t$_{min,1}$ (MJD) & 53482.251$\pm$0.003 \\
	a (R$_{\sun}$) & 23.3$\pm$0.8 \\
	i ($^{\circ}$) & 88.01$\pm$0.11 \\
	e              & 0.1303$\pm$0.0034 \\
	$\omega$ ($^{\circ}$) & 28.47$\pm$2.52 \\
	v$_{\gamma}$ (\kms) & 2.20$\pm$1.60 \\
	\end{tabular}

\end{multicols}
\endgroup
\tablecomments{Parameters and formal errors for PHOEBE fits to RV curves and lightcurves.}
\end{table*}
\clearpage

\end{document}